\documentclass[aps,twocolumn,prl,groupedaddress,showpacs]{revtex4}
\usepackage[T1]{fontenc}
\usepackage[latin9]{inputenc}
\usepackage{times}
\usepackage{color}
\usepackage{xspace}
\usepackage{amssymb,amsmath}
\usepackage{amsbsy}
\usepackage{graphicx}
\usepackage{multirow}
\usepackage{rotating}
\usepackage{microtype}
\usepackage{tabularx}
\usepackage{epstopdf}
\usepackage{float}
\usepackage{mathrsfs}
\usepackage{hyperref}
\usepackage[normalem]{ulem}
\def\be{\begin{equation}}
\def\ee{\end{equation}}
\def\e#1{\label{#1}\end{equation}}
\def\bea{\begin{eqnarray}}
\def\eea{\end{eqnarray}}

\def\ket#1{{|#1\rangle}}
\def\bra#1{{\langle#1|}}

\newcommand{\Niso}{$^{14}$N }
\begin{document}
\title{Non-classical measurement statistics induced by a coherent spin environment}
\author{D. D. Bhaktavatsala Rao $^1$}
\email{d.dasari@physik.uni-stuttgart.de}
\author{ Sen Yang $^2$}
\email{senyang@phy.cuhk.edu.hk}
\author{Stefan Jesenski $^1$}
\author{Florian Kaiser $^1$}
\author{J\"{o}rg Wrachtrup $^1$}

\affiliation{$^1$ 3. Physikalisches Institut, and MPI for Solid State Research, University of Stuttgart, Pfaffenwaldring 57, 70569 Stuttgart, Germany}

\affiliation{$^2$ Depatment of Physics, Chinese University of HongKong, Shatin, HongKong, China}

\date{\today}

\begin{abstract}
We demonstrate the role of measurement back-action of a coherent spin environment on the dynamics of a spin (qubit) coupled to it, by inducing non-classical (Quantum Random Walk like) statistics on its measurement trajectory. We show how the long-life time of the spin-bath allows it to correlate measurements of the qubit over many repetitions. We have used Nitrogen Vacancy centers in diamond as a model system, and the projective single-shot readout of the electron spin at low temperatures to simulate these effects. We show that the proposed theoretical model, explains the experimentally observed statistics and their application for quantum state engineering of spin ensembles towards desired states.
 \end{abstract}
\maketitle

{\bf Q}uantum mechanics allows us to post-select events that cannot be observed classically \cite{book1}. Even though these events are rare we cannot create a classical set-up that allows us to observe such an effect. Quantum random walk is one such example where the translation operation that can only displace the walker by a single step, in some (rare) cases can displace it much farther, while still keeping the average displacement the same \cite{QR1, QRev}. This comes about from the interference effect between the possible trajectories of a quantum random walker that would substantially modify the statistics observed in classical random walks. Such quantum walks are shown to be used to implement quantum search algorithms, and are also considered as a universal computational primitive \cite{QR2, QR3}. They have applications in the simulation of biological processes \cite{QR3}, and for implementing quantum algorithms \cite{QR4}, and they have been demonstrated on various experimental platforms \cite{exp1, exp2, exp3, exp4,exp5}. The discrete-time random walk can be described by the repeated application of a unitary evolution operator $U$ that acts on the combined space of a quantum coin and a quantum system with larger Hilbert space, a  multi-level system (e.g., graph, position or momentum space). The measurement result of the coin determines the transition of the walker between states of the multi-level system, similar to jumps on the nodes of a graph. Due to the quantum nature of the coin, the walk will also inhibit quantum features such as superposition, allowing for a fundamentally different random walk problem when compared to its classical counterpart \cite{QRrev}. It is known that in the case the result of the quantum coin is known, the quantum random walk reverts to classical random walk. Here we overcome this problem by showing that non-classical statistics could still be observed directly on the measurement statistics of the coin by exploiting the measurement induced back-action (memory) effects. Classically such memory effects of the walker have been considered in reinforced random walks \cite{RW} that are known to have broader applications in optimization problems but their quantum counterparts has not been explored

\begin{figure}
\includegraphics[width=0.5\textwidth]{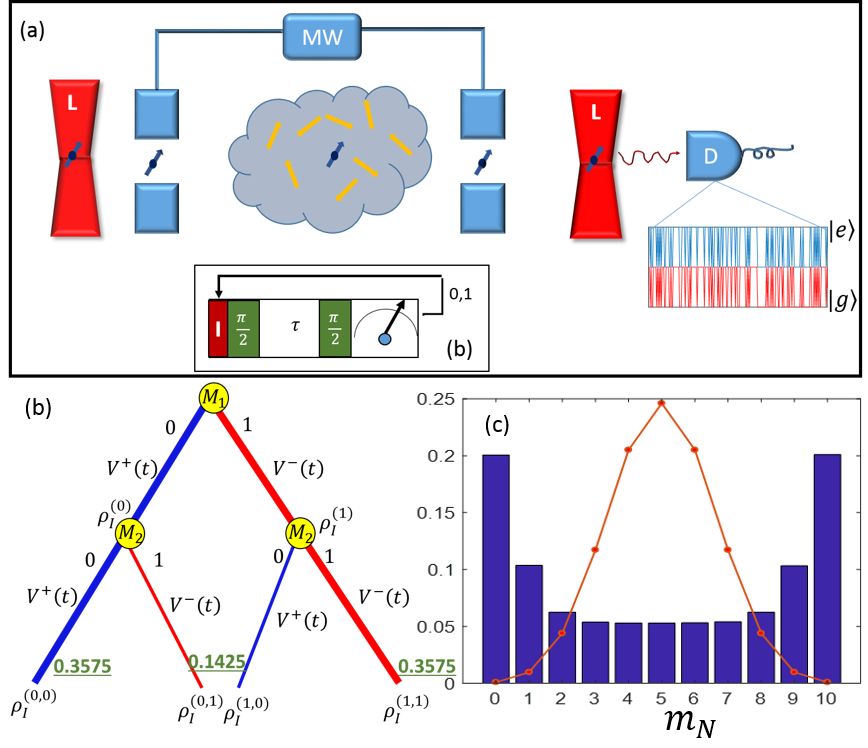}
	\caption{{(a) Schematic representation of the experimental scheme to test the non-classicality in measurements induced by a coherent spin environment. The NV spin is initialized-to and readout-from its ground state $\ket{0}$ optically. Microwave pulse (blue squares) are applied to create superposition of various spin states. The NV spin evolves by interacting with a spin environment for a contact time $\tau$, leading to phase evolution of its spin states. Jump statistics observed when the NV spin is projectively readout are shown in the inset. (b) The measurement result of the NV spin (coin) $M_n$ determines the evolution of the bath (walk space) by either of the superposition operators $V^\pm = (U^+_I \pm U^-_I)$, which is schematically depicted here. (c) The probability of occurrence of various measurements paths shown in (b) is plotted as a function of the path length $m_N$ which here represents the sum of bit values of each measurement string. We find Gaussian distribution (red solid-line) when all measurement paths are equally probable similar to CRW, while for walks conditioned on the bath state we find non-classical statistics mimicing a quantum random walk. }}
\end{figure}

Open quantum systems can naturally display non-Markovianity (memory effects) in their dynamics when interacting with environments having long correlation times \cite{book2}. In our case the reinforcement arises with each measurement collapsing the environment to a smaller subspace of states thereby modifying the space (graph) on which the further walk takes place. With repeated measurements yielding similar result, this subspace eventually shrinks to a single quantum state indicating steady state dynamics \cite{zenocool,ddb1,ddb2}.

\begin{figure*}
\includegraphics[width=0.98\textwidth]{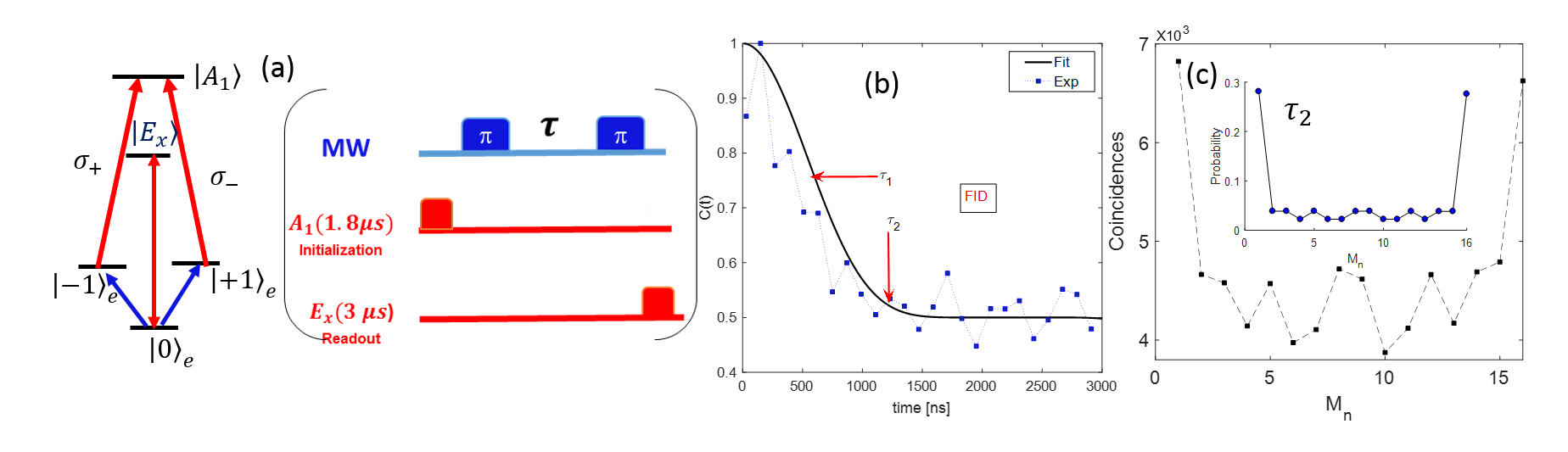}
	\caption{\textbf{Non-classical measurement statistics} (a) Schematics of the level-diagram and the pulse sequence used for experiments is shown. (b) The free induction decay of the coherence of the NV spin $C(t)$ is plotted as a function of time $t$. (c) The coincidences for finding a given measurement string $M_n$ in four consecutive measurements is plotted when measurements are performed at intervals of $\tau_2$. In the inset we show measurement statistics obtained from theoretical simulation of dynamics governed by Eqs (3, 4).  }
\end{figure*}

Our model consists of a single NV center consisting of an electronic spin (S=1) and intrinsic \Niso nuclear spin (I=1), coupled by hyperfine interaction \cite{nvrev}. In addition to this intrinsic nuclear spin, the NV center is immersed in a spin-bath of ${}^{13}C$ nuclear spins and the electronic spins of the P1 centers in diamond \cite{ref6}. Due to a very weak dipolar coupling among the nuclear spins, that dominantly constitute the spin-bath we consider it noninteracting.  The dipolar coupling between the spin bath and the NV center is determined by the quantization axis of the NV center, rendering all components of the dipolar coupling non-zero. In the reference frame of the central spin (NV) one can think of the spins of the bath as being randomly distributed in a three-dimensional plane comprising of the $S$-spin polarization and the plane perpendicular to it. 
With the magnetic field aligned along the central spin axis (say NV $z$-axis) the Hamiltonian that determines the dynamics is given by\cite{nvrev,ddb2} 
\begin{equation}
H = S^z \otimes \sum_k \vec{g}_k (r) \cdot \vec{I}_k + \omega \sum_k I \otimes I_z^{(k)}
\label{eq:totalHamiltonian}
\end{equation}
where $g_k$ is the strength of dipolar coupling between the NV (central) spin and the $k$-th nuclear spin, dependent on the spatial separation between these spins. The random spatial location of the spins with respect to the NVC leads to inhomogeneous couplings such that the bath has neither conserved quantities nor preferred symmetries. The external field $\omega$, (the nuclear Zeeman term) under which the nuclear spins precess is assumed to be uniform over the entire sample and is along the $z$-direction. The dynamics generated by the above Hamiltonian can been exactly solved \cite{ddb3}. 
In the $S^z$ basis of the central spin ($|{1}\rangle, |-1\rangle$), the above Hamiltonian can be rewritten as 
\bea
H &=& H^+\ket{+1}\bra{+1} + H^-\ket{-1}\bra{-1}, \nonumber \\
H^{\pm} &=& \sum_k \omega_k I^k_z \pm \sum_k \vec{g}_k (r) \cdot \vec{I}_k
\label{eq:HPlusMinus} 
\eea

where $H^{\pm}$ are the nuclear spin-bath operators. 
The simple form of $H^{\pm}$ makes it easy to diagonalize, so that we obtain a closed-form equation for the time-evolution operator of the total system
\be
\label{uop}
U(t) =\left[U^+_I(t)|1\rangle\langle 1| \, + \, U^-_I(t)|{ -1}\rangle\langle-1|\right]
\ee
where $U^\pm_I(t)$ govern the dynamics of the spin-environment conditioned on the state of the NV spin \cite{supp}.

Here one can make the analogy to the quantum random walk (QRW) problem, where the evolution in the walker space (spin-bath) is conditioned on the state of the quantum coin (NV spin) through Eq. (3). As opposed to the known QRW problem, we do not preserve coherences among various paths as the walker's state is readout after each step \cite{QRev}. Equally we cannot readout the walker's state as it would require a full quantum tomography of the spin-bath, which is technically quite demanding \cite{plenio}. We instead, infer the walk statistics through the measurement statistics of the NV-spin itself by taking advantage of the long-life time of the bath spins, which allows it to store the measurement induced back-action so as to influence future measurements. As the coin operation is strongly dependent on the current state of the walker (due to the nature of interaction between them), we find that as the walker moves from his original position, the probability distribution of the coin tosses change accordingly biasing the walk similar to reinforced random walks discussed earlier. The resultant statistics mimicking QRW arises due to the measurement  back-action effects where the measurement influences the bath state, which in turn influences the succeeding measurement (see Suppl. for details). 

In the absence of external field the Hamiltonian describing the system-bath interaction takes a simpler form $H = S^z \otimes \hat{B}$ that could be mimicked by a classical noise model, where the bath operator $\hat{B} = \sum_k \vec{g}_k \cdot \vec{I}$. By considering the initial bath state to be completely unpolarized the bath operator can be replaced by a random magnetic field to obtain similar dynamics on the TLS. Starting from an initial superposition state, $\ket{\psi} = \frac{1}{\sqrt{2}}[\ket{0}+\ket{1}]$, the coherence of the TLS for any later time $t$, is given by
\be
C(t)\equiv \langle S^x \rangle = \int^{b_m}_{-b_m}d\omega G(\omega) \cos (\omega t),
\ee
where $b_m$ are all possible energy eigenstates of the bath (see Supp.) and $G$ is the bath distribution function.  Due to the dynamics governed by Eq. (), the coherence is lost at a rate $\gamma$ determined by the width of the distribution $G(\omega)$. This sets the $T_2^*$ time of the spin, and does not change for a given spin environment. On the other hand, the dephasing time $T_2$ could be made longer by employing dynamical decoupling sequences. The simpler form of the interaction given above does not lead to any bath dynamics, i.e., $\rho_B(t) = \rho_B(0)$, when the initial bath state is completely mixed. For this reason, the bath acts like a background noise on the TLS and almost resembles a classical noise source for the TLS.  The changes in the bath state could result from the measurement back action inducing a non-unitary action on its state through operators $V_\pm(t) = (U^+_I(t)\pm U^-_I(t))/2$ depending on the measurement result $'0'$ or $'1'$. For example, a measurement outcome that finds the TLS in state $0$ all the $n$-times, will give rise to a measurement string $M_1 = \lbrace{0,0,\cdots 0 \rbrace}$, and the bath state corresponding to such a measurement trajectory is given by \cite{ddb1}
\be
\rho^{(1)}_B(t) = \frac{V_+\rho_B(0)V_+^\dagger}{{\rm Tr}[V_+\rho_B(0)V_+^\dagger]}.
\ee
where $V = \cos^n (\hat{B}t)$. Clearly $\rho^{(1)}_B(t) \ne \rho_B(0)$, indicating the modified bath state for a given trajectory $M_1$. Similarly all bath states corresponding to the $2^n$ measurements can found. Interestingly, one can verify that the total averaged state equals to the actual state at time $t$ i.e., $\sum_M \rho^{(M)}_B(t) = \rho_B(t)$, confirming the {\it ergodicity} of system when averaged over all possible measurements (see Suppl. for details). The hierarchy of changes caused by the bath on the system and by the system back on the bath will continues until a steady state is reached, that has comparatively larger polarization than its initial state. This polarization gain adds also on to the measurement statistics, explaining the preferred choice of measurement statistics as we discuss below.

To demonstrate these effects we implement this protocol in a low strain ($\approx$1.2 GHz) NV center to suppress strain-induced effects, e.g. lowering the symmetry of the NV and altering the configuration of the excited state \cite{nvrev}. For experiments a NV center along [111] orientation is chosen \cite{yang}. In the experiment the applied magnetic field is oriented along the NV-axis, and we have considered two cases: (i) the net field is zero, rendering $\ket{\pm 1}_e$ degnerate and (ii) a non-zero field that lifts the degeneracy between these spin states. To ensure projective readout of the NV spin state, the experiments are performed at low temperatures (4K), at which the optical selection rules allows one to distinguish the electronic spin states $\ket{\pm 1}_e$ and $\ket{\pm 0}_e$  with $ > 99\%$ fidelity through single shot readout \cite{ref8}.
The upper limit of the nuclear spin coherence time, given by the electron spin's $T_1$ time which reaches minutes at low temperatures, allowing for larger number of steps in QRW protocol. Various steps in the protocol are schematically shown in Fig.1, where we initialize the electron spin in state $\ket{0}$ by resonantly exciting it on the $A_1$ transition, followed by a microwave transition: $\ket{0} \rightarrow [\ket{+1} + \ket{-1}]/\sqrt{2}$, and further its phase evolution through interaction with the spin-bath. Finally another microwave transition and then by resonantly exciting the center on the $E_y$ transition, we projectively readout the NV center i.e., we observe florescence if it is  in the state $\ket{0}$, and remains dark if it is in the other states. There is a slight probability for the spin to end up in a state orthogonal to its projective basis with a small probability of $0.01$ giving rise to errors in the observed measurement statistics \cite{Supp}.

We first measure the decay of NV spin coherence (FID) due to bath interaction (see Fig. 2(a)), and we find that  NV spin is completely decohered over a time scale of for $T_2^* \sim 1.2\mu$s. We now perform four measurements (and repeat them $75312$ times for statistics) which will result in $2^4 = 16$ possible measurement strings (represented as basis states of a 4-qubit system in '0', '1' basis), and their occurrence probabilities will determine the non-classicality in their statistics and the coherent nature of the spin-bath. The choice of interaction $\tau$ is set from the FID behavior shown in Fig. 2. To mimic the classical coin, we chooose the time $\tau =T^*_2 = 1.2 \mu$s, such that the NV spin state is completely has become depolarized, and the probability of finding it either of its orthogonal state is equal to $0.5$, representing a classical coin. If the initial conditions are similar in every experimental run, all the $16$ possibilities should occur with equal probability. On the contrary, for the statistics shown in Fig. 2(b), there is a high probability for measurement results that are identical i.e., $M_1 = 0000$ and $M_{16} = 1111$, and decreases with increasing inhomogeneity. This is due to the fact that though we reset the NV spin after each measurement, the long coherence (life-time) of the nuclear spin-bath does not reset itself to the initial fully mixed state (as $\tau \ll T_{1n}$), rather gets projected to a different state after the each measurement effecting the later measurement. Thus a different bath state at each measurement, modifies the measurement. 

We further observe symmetries in the statistics i.e., measurements with unequal number of $0$ and $1$'s occur with same probability i.e., strings $0001$, $0100$, $1000$, $0100$ occur with equal probability. To get a deeper insight into the physical picture of these measurement statistics we represent all possible measurement strings as a basis set ($2^N$) of $N$-qubit system. The basis can be divided into subspaces with conserved total $z$-component i.e., sum of all bit values. Thus there are $N$ subspaces whose dimension ${}^NC_{N/2-m}$ ($0 \le m \le N$) displays a Gaussian distribution $~\exp(-m^2/N)$. If all the measurements strings are equally probable similar to CRW, then the measurement statistics follow a Gaussian distribution with a width $\sqrt{N}$ and centered at $m = N/2$. On the otherhand we experimentally find a different distribution, where the maximally occurring strings are the extreme ends i.e., at $m=0$ ($M_n = '0000'$) and $m=16$ ($M_n = '1111'$), and decreasing probabilities equally from either side. Such measurement statistics mimics the QRW behavior shown earlier \cite{QRev}.
%


\begin{figure}
\includegraphics[width=0.5\textwidth]{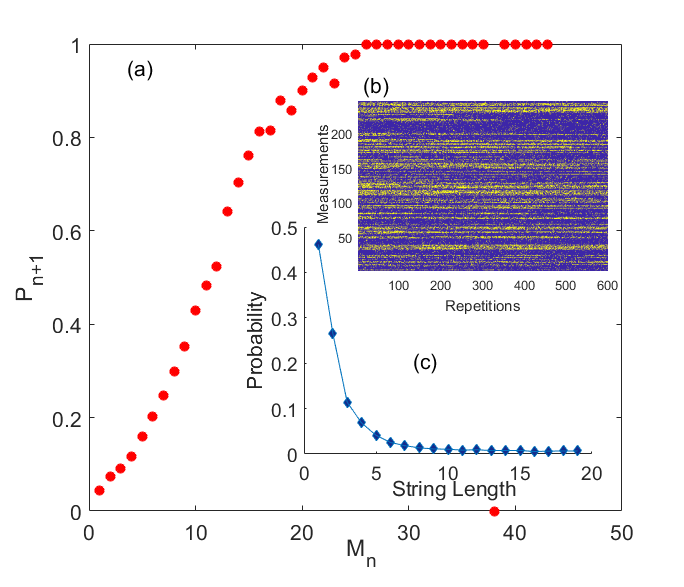}
	\caption{\textbf{(a)} Probability of finding a similar measurement result after $n$ identical measurement results is shown as a function of $n$ in a measurement string $M_n$, obtained from the data shown in (b). \textbf{(b)} Raw data showing the results of $246$ measurements repeated $600$ times. Violet indicates result $0$, and yellow result $1$.  \textbf{(c)} The probability of finding a measurement string of length $n$ (which is the sum of bit values of measurement results for a given length of the string) with identical results obtained from (b). }
\end{figure} 

If the observed FID measurement result at a given time $\tau$ is considered to be a macroscopic variable, then the total number of measurements (strings) and their distribution will determine the microscopic states of the system, making a nice analogy to statistical physics and thermodynamics. While the average of the statistics agrees with the observed FID results at those respective times, the probability with which different measurements are chosen adds on to them non-classical statistics induced by measurement back-action. For example, if the measurement results are all equally probable, then such a distribution (when written in the computational basis of four qubit system) has maximum entropy of $1/16$ \cite{book3}, and the FID result corresponding to such statistics is a fully depolarized NV spin as shown in Fig. 2(a) at time $\tau=1.2 \mu$s. On the other hand the distribution shown in Fig. 2(b) gives a similar FID result but the entropy of the observed measurements is lower, $2.3/16$, indicating a preferential ordering or polarization in measurement statistics. To understand the origin of this entropy loss (purity), we will evaluate the bath dynamics governed by Eq. (5). For example, if we consider a bath consisting of four spins, that are coupled to the NV spin resulting in a similar FID behavior as observed experimentally in Fig. 2(a), and then by performing four measurements, at time $\tau$ i.e., when the spin is completely depolarized, we find statistics showing similar behavior (see inset of Fig. 2(b)). The average purity of the four spin-state after observing the $16$ possible measurement strings is found to be $\sim 2.1/16$, which is quite close to the measurement purity ($ \sim 1. 7/16$) obtained form the measurement statistics shown in the inset of 2(b). The measurement back-to-back action on the system-bath dynamics thus reveals the physical nature of the bath that cannot be mimicked by classical noise models \cite{book2}.

To further confirm the correlations in subsequent measurements seen in Fig. 2, we have performed $n=246$ measurements and repeated the same $600$ time, to obtain statistics shown in Fig. 3 (b). Since the measurement influences the bath-state, finding identical result in $m$ consecutive measurement steps will influence the $(m + 1)-th$ measurement, and also in turn confirms the stabilization of the bath state. We see this increasing probability and saturation in our measurements which could hint a steady state for the bath with higher polarization. 

In conclusion we have shown that when measuring a quantum system coupled to a coherent spin environment, the measurement back-action modifies the environment, which in turn influences the subsequent measurement, thus generating correlations among them. The microscopic detail on the occurrence of various measurement strings helps to understand the non-classical features in quantum measurements and system-bath interactions, allowing for the purification (cooling) of a quantum environment into desired target states\cite{ddb1,ddb2}. By finding the measurement distributions that result in observable average behavior of system, we will be able to understand the statistical nature of measurements and further to their analogy with thermodynamics \cite{ren-bao,noam}.

\begin{acknowledgements}
        We would like to acknowledge the financial support by the ERC project SQUTEC, DFG (FOR1693), DFG SFB/TR21, EU DIADEMS, SIQS, Max Planck Society and the Volkswagenstiftung as well as the Baden-Wuerttemberg Foundation. SY would like to acknowledge the support from Hong Kong RGC/ECS 24304617.
\end{acknowledgements}

\end{document}